\documentclass[
    aps,
    prx,
    superscriptaddress,
    twocolumn,
    a4paper,
    10pt,
    floatfix,
    longbibliography
]{revtex4-2}

\usepackage[american]{babel}

\usepackage{grffile}

\usepackage{newtxtext,newtxmath}
\usepackage{microtype}

\usepackage{amsmath}
\usepackage{amssymb}
\usepackage{bbm}

\usepackage{mathtools}
\usepackage{dsfont}
\usepackage{braket}
\usepackage{cancel}
\usepackage{slashed}

\usepackage{graphicx}
\usepackage[svgnames]{xcolor}

\usepackage{siunitx}

\usepackage{ragged2e}
\usepackage{array}
\usepackage{tabularx}
\usepackage{booktabs}

\definecolor{cset-aps-blueberry}{RGB}{28,128,158}
\definecolor{cset-aps-blue}{RGB}{46,44,184}
\definecolor{cset-aps-turquoise}{RGB}{0,67,88}
\definecolor{cset-aps-limegreen}{RGB}{190,219,67}
\definecolor{cset-aps-green}{RGB}{31,138,112}
\definecolor{cset-aps-yellow}{RGB}{255,225,25}
\definecolor{cset-aps-orange}{RGB}{253,116,0}
\definecolor{cset-aps-red}{RGB}{219,0,43}

\usepackage{tikz}

\usepackage{pgfplots}
\pgfplotsset{%
    every axis legend/.append style={%
        cells={anchor=west},
        at={(0.96,0.04)},
        anchor=south east,
        font=\scriptsize,
        },
    every axis/.append style={%
        yticklabel style={%
            /pgf/number format/fixed zerofill,
            /pgf/number format/precision=2},
        },
    width= \textwidth,
    height=8cm,
    xmajorgrids=true,
    xminorgrids=false,
    minor x tick num=1,
}
\usepgfplotslibrary{external}

\usetikzlibrary{decorations}

\usepackage{pict2e,picture}

\makeatletter
\DeclareRobustCommand{\Arrow}[1][]{%
\check@mathfonts
\if\relax\detokenize{#1}\relax
\settowidth{\dimen@}{$\m@th\rightarrow$}%
\else
\setlength{\dimen@}{#1}%
\fi
\sbox\z@{\usefont{U}{lasy}{m}{n}\symbol{41}}%
\begin{picture}(\dimen@,\ht\z@)
\roundcap
\put(\dimexpr\dimen@-.7\wd\z@,0){\usebox\z@}
\put(0,\fontdimen22\textfont2){\line(1,0){\dimen@}}
\end{picture}%
}
\makeatother

\usepackage{hyperref}
\hypersetup{%
    colorlinks=true,
    linkcolor={cset-aps-red},
    linkbordercolor={cset-aps-red},
    filecolor={cset-aps-orange},
    filebordercolor={cset-aps-orange},
    citecolor={cset-aps-blue},
    citebordercolor={cset-aps-blue},
    urlcolor={cset-aps-green},
    urlbordercolor={cset-aps-green},
    menucolor={cset-aps-limegreen},
    menubordercolor={cset-aps-limegreen},
    breaklinks=true,
    pdfborderstyle={/S/U/W 2},
    pdfpagemode=UseOutlines,
    pdfstartpage={1},
}

\newcommand{\ee}{\textsf{e}}
\newcommand{\ii}{\textsf{i}}

\usepackage{bm}
\renewcommand{\vec}[1]{\bm{#1}}

\newcommand{\orcid}[1]{\href{https://orcid.org/#1}{\includegraphics[width=7pt]{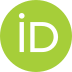}}}

\usepackage{comment}

\usepackage{lipsum}

\usepackage[normalem]{ulem}

\pgfplotsset{compat=1.16}

\begin{document}

\title[General relativistic center-of-mass coordinates for composite quantum particles]{General relativistic center-of-mass coordinates for composite quantum particles}
\collaboration{This article has been published in \href{https://doi.org/10.1103/PhysRevD.111.064005}{Phys. Rev. D \textbf{111}, 064005 (2025)}.}
\author{Gregor Janson~\orcid{0009-0002-7509-4550},}
\email{gregor.janson@uni-ulm.de}
\author{Richard Lopp}
\affiliation{Institut f{\"u}r Quantenphysik and Center for Integrated Quantum
    Science and Technology (IQST), Universit{\"a}t Ulm, Albert-Einstein-Allee 11, D-89069 Ulm, Germany}

\begin{abstract}
\noindent
Recent proposals suggested quantum clock interferometry for tests of the Einstein equivalence principle. However, atom interferometric models often include relativistic effects only in an \textit{ad hoc} fashion. Here, instead, we start from the multiparticle nature of quantum-delocalizable atoms in curved spacetime and generalize the special-relativistic center of mass (c.m.) and relative coordinates that have previously been studied for Minkowski spacetime to obtain the light-matter dynamics in curved spacetime.
In particular, for a local Schwarzschild observer located at the surface of the Earth using Fermi-Walker coordinates, we find gravitational correction terms for the Poincaré symmetry generators and use them to derive general relativistic c.m. and relative coordinates. In these coordinates we obtain the Hamiltonian of a fully first-quantized two-particle atom interacting with the electromagnetic field in curved spacetime that naturally incorporates special and general relativistic effects.
\end{abstract}

\maketitle

\section{Introduction}
\label{sec.Introduction}
Light-pulse atom interferometers are high precision instruments that have demonstrated their effectiveness in a wide range of applications, including measurements of gravitational acceleration~\cite{Kasevich1991,Peters1998,Peters1999}, rotation~\cite{Lan2012}, and Newton's gravitational constant~\cite{Rosi2014}, as well as in field applications~\cite{Barrett2016,Wu2019,Templier2022,Stray2022} and mobile gravimetry~\cite{Wu2019}. The most accurate determination of the fine structure constant to date was achieved using atom interferometry~\cite{Parker2018,Morel2020}. Proposals for interferometer schemes have also been put forward to test the universality of gravitational redshift~\cite{Roura2020,Ufrecht2020} and the universality of free fall~\cite{Ufrecht2020}, as well as gravitational wave detection~\cite{Dimopoulos2008b,Dimopoulos2009,Graham2016}. Prototypes, whose construction recently began, could be sensitive to ultra-light dark matter signals~\cite{Geraci2016,Arvanitaki2018,Badurina2023a}, making them promising candidates as testbeds for gravitational wave antennas~\cite{Abe2021,Bertoldi2021,Badurina2021} based on atom interferometry.

These applications show that there is a lot of interest in gravitational measurements via atom interferometry. Most theoretical descriptions are based on an \textit{ad hoc} addition of special-relativistic and gravitational effects such as the mass defect and gravitational potentials, c.f., Refs.~\cite{Zych2011,Zych2017,Loriani2019,Ufrecht2020,Roura2020,DiPumpo2021,DiPumpo2022,DiPumpo2023,Janson2024}. First approaches to include general relativistic (GR) effects in atom interferometer phases from first principles have already be carried out by Dimopoulos et al.~\cite{Dimopoulos2008} and Werner et al.~\cite{Werner2024}. A post-Newtonian description of a two-particle atom (e.g. a proton and an electron) in weakly curved spacetime -- including external electromagnetic fields -- has been given by Schwartz and Giulini~\cite{Schwartz2019,Schwartz2020b}, which is an extension of the work of Sonnleitner et al.~\cite{Sonnleitner2018} to curved spacetimes. Sonnleitner et al.~\cite{Sonnleitner2018} showed in their paper that by using special-relativistic corrected center-of-mass (c.m.) and relative coordinates~\cite{Osborn1968,Close1970} they can decouple the dynamics of internal and external degrees of freedom in such a way that the remaining cross-terms can be interpreted as the mass defect, i.e., the total mass of the atom depends on the internal state. Schwartz and Giulini~\cite{Schwartz2019,Schwartz2020b}, however, found an additional cross-term between the internal and external dynamics induced by the gravitational field that does not vanish using these special-relativistic c.m. and relative coordinates.
There are, though, two issues that one needs to consider for a proper treatment of the matter. 

\paragraph{Local observer} The starting point of Schwartz and Giulini in Refs.~\cite{Schwartz2019,Schwartz2020b} is the Eddington-Robertson parametrised post-Newtonian metric, c.f. equation (2.1) in Ref.~\cite{Schwartz2019} and (2.5.1) in Ref.~\cite{Schwartz2020b}.
This metric is an excellent starting point for examining deviations from GR.
However, it describes physical phenomena as seen by a static observer at infinite distance of the sourcing mass, e.g., the Earth.
For an accurate description of physical phenomena as seen by a realistic observer, we first have to describe everything within a proper reference frame, i.e., the Fermi-Walker frame~\cite{Nesterov1999,Manasse2004,Klein2008,Klein2009,Poisson2011,Klein2012,Martin-Martinez2020,Perche2021,Perche2022}.
Fermi-Walker coordinates (FWC) are characterized by a tetrad basis attached to an (accelerated) observer, where the timelike basis vector is parallel to the four-velocity of the observer's worldline and the spatial triad does not rotate.
They have been used, e.g., by Perche et al. in Refs.~\cite{Perche2021,Perche2022} to describe a static atom with the nucleus of the atom located at the origin of the Fermi-Walker frame, which is tied to the position of the (possibly accelerated) observer.
However, this approach does not account for the full dynamical two-body nature of the system leading to c.m. and relative dynamics of the composite particle.
In this paper we will restrict ourselves to non-rotating observers. Generalizations would employ  \textit{generalized} FWC~\cite{Kajari2009,Kajari2011,Llosa2017}, which have been used, e.g., in the context of relativistic Sagnac interferometry by Kajari et al. in Refs.~\cite{Kajari2009,Kajari2011}.

\paragraph{Gravitationally corrected degrees of freedom} Already in the 1950s the relativistic dynamics of systems of multiple particles aroused the interest of physicists:
The dynamics of systems of particles with and without internal interactions have been shown to be relativistically invariant when the condition of invariant world lines is dropped~\cite{Thomas1952}.
Further, the separability of internal and external degrees of freedom has been investigated for non-interacting particles~\cite{Dirac1949,Macfarlane1963}. Internal interactions were inserted into the rest mass by Bakamjian et al.~\cite{Bakamjian1953}, c.f., the mass defect -- one of the aforementioned \textit{ad hoc} insertions. 
Afterwards, Foldy~\cite{Foldy1961} confirmed the results of Bakamjian et al.~\cite{Bakamjian1953} by considering a system of a fixed number of directly interacting particles.
When separating the internal and external dynamics, one usually introduces c.m. and relative coordinates and momenta. When dealing with relativistic particles, these coordinates known from undergraduate textbooks are relativistically corrected, as shown by Osborn and Close for systems of vanishing total electric charge~\cite{Osborn1968,Close1970}. These relativistically corrected coordinates have shown to be extremely useful to decouple the internal and external degrees of freedom of a two-particle atom~\cite{Sonnleitner2018}. Martínez-Lahuerta et al. found a generalization of these coordinates for systems of non-vanishing total charge~\cite{Martinez-Lahuerta2022}.
While Osborn and Close~\cite{Osborn1968,Close1970} found the relativistic internal coordinates via a singular Gartenhaus-Schwartz transformation, Liou found a more elegant way to determine the relative coordinates via a unitary transformation~\cite{Liou1974}.
Ultimately, there is the work of Krajcik and Foldy where special-relativistic c.m. and relative coordinates are calculated for composite systems with arbitrary internal interactions~\cite{Krajcik1974}.
All these have one thing in common: they make use of the generators of the Poincaré group, i.e., the symmetry group of flat spacetime. In the case of curved spacetime, however, the symmetry generators do not look the same as in flat spacetime. Consequently, the relativistic c.m. and relative coordinates of Refs.~\cite{Osborn1968,Close1970,Krajcik1974,Liou1974,Martinez-Lahuerta2022} should be modified when dealing with gravitation. The goal of this paper is to find these GR corrections for weakly curved spacetimes and use them to obtain a Hamiltonian describing a two-particle atom interacting with external electromagnetic fields, in analogy to Schwartz and Giulini~\cite{Schwartz2019,Schwartz2020b}, yet for a realistic local observer on Earth.

\subsection*{Overview \& Structure}
In the following we shortly describe the structure of this paper:
In Sec.~\ref{Sec: Metric in FWC} we recapitulate the basics of FWC and use them to expand the Schwarzschild metric around the worldline of a static observer located at the equator of the Earth up to second order in FWC.
In Sec.~\ref{Sec: Generators of Symmetry Group} we use this metric to find gravitational correction terms for the well-known Poincaré symmetry generators up to first order in $\epsilon=R_S/R_E$ via the Killing equations, where $R_S$ is the Schwarzschild radius and $R_E$ is the radius of the Earth.
With these corrected symmetry generators in hand, we can then calculate the GR corrections for the c.m. and relative coordinates in Sec.~\ref{Sec: Generalized c.m. and relative Coordinates}.
Finally, in Sec.~\ref{Sec: Two-Particle Dynamics in Curved Spacetime} we use the metric of Sec.~\ref{Sec: Metric in FWC} and the generalized c.m. and relative coordinates of Sec.~\ref{Sec: Generalized c.m. and relative Coordinates} to calculate -- in analogy to Schwartz and Giulini~\cite{Schwartz2019,Schwartz2020b} -- the full Hamiltonian describing a two-particle atom with vanishing total charge in curved spacetime including external electromagnetic fields.
We then conclude with a summary and a discussion of our results in Sec.~\ref{Sec: Conclusion}.

\section{Metric in Fermi-Walker frame}
\label{Sec: Metric in FWC}
In order to obtain physical observables corresponding to an experiment witnessed by an observer, we have to describe the dynamics in the proper coordinate frame~\cite{Nesterov1999,Manasse2004,Klein2008,Klein2009,Poisson2011,Klein2012,Martin-Martinez2020,Perche2021,Perche2022,Kajari2009,Kajari2011,Llosa2017}. Let us assume that we have a four-dimensional manifold $\mathcal{M}$ with a Lorentzian  metric $g^\prime_{\mu\nu}(y^\alpha)$ given in \textit{a priori} coordinates $y^\alpha$. Subsequently this will be chosen to be the Schwarzschild metric given in Schwarzschild coordinates. These coordinates, however, do not describe physical phenomena as seen by a realistic observer. Instead, we have to find a coordinate system attached to the worldline $\sigma^\alpha(\tau)$ of the observer, parametrized by its proper time $\tau$. For this we choose the FWC $x^\alpha$. The goal of this section is to provide a short introduction to the Fermi-Walker frame which will serve as the basis for the generalized c.m. and relative coordinates in curved spacetime.

\begin{figure}
    \centering
    \includegraphics[width=0.5\linewidth]{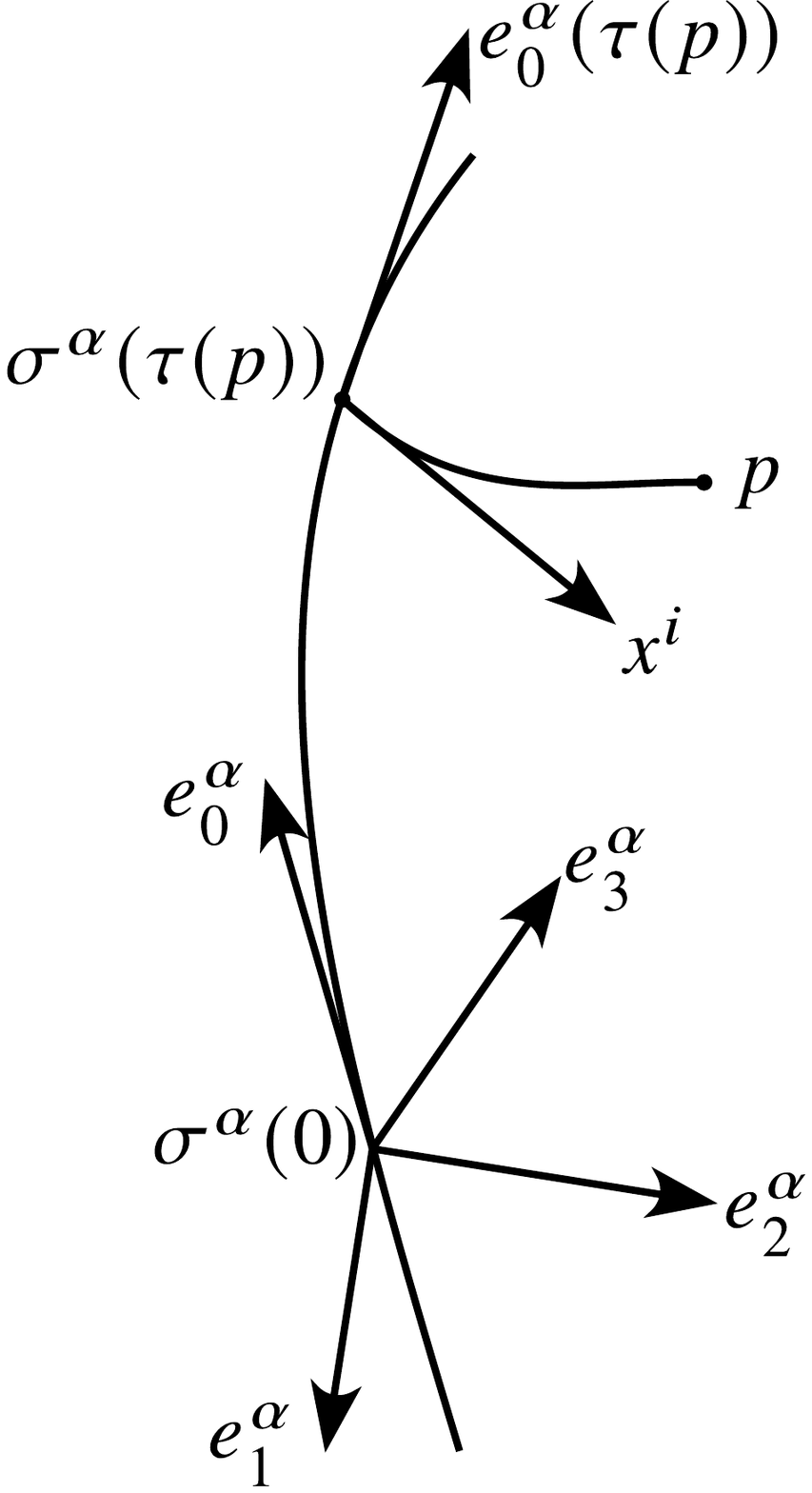}
    \caption{Orthonormal tetrad $e_{(\mu)}^\alpha(\tau)$ Fermi-Walker transported along the worldline $\sigma^\alpha(\tau)$ of the observer.}
    \label{fig: FWC}
\end{figure}

To do so we firstly have to define the orthonormal tetrad basis $e_{(\alpha)}^\mu(\tau)$ attached to the observer. Note, that the indices in brackets are tetrad indices in contrast to coordinate indices. They are raised and lowered by the Minkowski metric \mbox{$\eta=\mathrm{diag}(-1,1,1,1)$}. The time coordinate in FWC is given by the proper time, i.e., $x^0 = \tau$. Let us define the basis at some given time $\tau_0 = 0$. We can then choose the timelike tetrad to be the tangent vector to the worldline: $e_{(0)}^\mu(0)  = \mathrm{d}\sigma^\mu/\mathrm{d}\tau (0)=u^\mu(0)$. The remaining three spatial tetrads can be chosen accordingly so that they build a right-handed basis for the spatial tangential space and fulfill the orthonormality condition
\begin{align}\label{eq: orthonormality condition}
    g^\prime_{\mu\nu}e^\mu_{(\alpha)}(0)e^\nu_{(\beta)}(0) = \eta_{(\alpha\beta)}.
\end{align}
To take into account the motion of the observer in curved spacetime, the tetrads have to be correctly transferred from one tangential space to another. That is, they have to be Fermi-Walker transported along the worldline $\sigma^\alpha(\tau)$
\begin{align}\label{eq: Fermi Walker transport}
    e_{(\alpha);\nu}^{\mu}u^\nu - \frac{1}{c^2}\left(u^\mu a_\beta - a^\mu u_\beta\right) e_{(\alpha)}^\beta= 0,
\end{align}
where $a_\beta$ is the four-acceleration.
Fermi-Walker transport ensures an instantaneous rest frame that is non-rotating under the observer's motion, and reduces to parallel transport for geodesics.
Furthermore, it follows that the tetrads continue to be orthonormal at all times $\tau$. It is easy to see that the four-velocity $u^\alpha$ automatically fulfills the transport equation, Eq.~\eqref{eq: Fermi Walker transport}, so that the timelike tetrad can be chosen to be $e_{(0)}^\alpha(\tau)=u^\alpha(\tau)$ for all times $\tau$, c.f., Fig.~\ref{fig: FWC}.

For the observer's wordline we define the normal neighborhood $\mathcal{N}_{\sigma^\alpha(\tau)}$ such that all points in that neighborhood can be connected to $\sigma^\alpha(\tau)$ via a unique geodesic. At each time $\tau$ we can then define the rest surface $\Sigma_\tau \subset \mathcal{N}_{\sigma^\alpha(\tau)}$ spanned by all geodesics that are orthogonal to $u^\alpha(\tau)$ at $\sigma^\alpha(\tau)$, i.e., a local foliation of spacetime around $\sigma^\alpha(\tau)$. To describe a point $p$ via FWC, we firstly have to find the rest surface $\Sigma_\tau$ that contains $p$ and find the corresponding time $\tau$. After that we can assign the spatial coordinates $x^i$ so that
\begin{equation} 
    p = \exp_{\sigma^\alpha(\tau)}(x^i e_{(i)}(\tau)),
\end{equation}
where $\exp_{\sigma^\alpha(\tau)}$ is the exponential map at the point $\sigma^\alpha(\tau)$.
In this Fermi-Walker frame, the spatial distance between a point $x^\alpha$ and the observer is then given by $r=\sqrt{\delta_{ij}x^ix^j}$.

Finally, we can write down the metric $g_{\mu\nu}(x^\alpha)$ in FWC via an expansion in $r$ to second order, i.e.,
\begin{subequations}\label{eq: Metric in FWC (general)}
\begin{align}
        g_{00} &= -(1+2a_i(\tau)x^i+(a_i(\tau)x^i)^2)+R_{0i0j}(\tau)x^i x^j + \mathcal{O}(r^3),\\
        g_{0i} &= -\frac{2}{3}R_{0kij}(\tau)x^k x^j + \mathcal{O}(r^3),\\
        g_{ij} &= \delta_{ij} - \frac{1}{3}R_{ikjl}(\tau) x^k x^l + \mathcal{O}(r^3),
\end{align}
\end{subequations}
where $a^\mu(\tau) = u^\mu_{;\nu}(\tau)u^\nu(\tau)$ is the four-acceleration of $\sigma^\alpha(\tau)$ and $R_{\alpha\beta\gamma\delta}(\tau)$ is the Riemann curvature tensor on $\sigma^\alpha(\tau)$ in the new Fermi-Walker basis~\cite{Nesterov1999,Kajari2011,Martin-Martinez2020,Perche2021,Perche2022}. The four-acceleration $a^\mu$ and the Riemannian curvature tensor $R_{\alpha\beta\gamma\delta}$ can be easily calculated within the \textit{a priori }basis. In order to express them in the new coordinates, we need to determine the Jacobian of the coordinate transformation from \textit{a priori }to FWC, evaluated on the worldline only, i.e.,~\cite{Klein2008} 
\begin{align}\label{eq: Jacobian on worldline}
    J^\alpha_\beta\vert_\sigma = \frac{\partial y^\alpha}{\partial x^\beta}\vert_\sigma = e^\alpha_{(\beta)}.
\end{align}

\subsection{Static Observer in Schwarzschild Spacetime}
Let us now particularize our setting, without loss of generality, to the case of an observer on the Earth's equator. We may model this via a static observer in a Schwarzschild spacetime, provided the experiments take place over a sufficiently small time period compared to the time scale associated with the angular frequency of the rotating Earth, i.e. \mbox{$\omega_\textsf{Earth}\approx 7.29\times10^{-5}\mathrm{s}^{-1}$}. Therefore, the Schwarzschild metric is sufficiently suited to obtain first order gravitational effects in atomic experiments. However, one could also apply our method to, e.g., the Kerr metric. In those general cases, however, one would obtain semi-analytical results in contrast to the analytical results for the Schwarzschild scenario in the following.

In detail, let us consider the Schwarzschild metric that belongs to a static, spherically symmetric mass distribution in Schwarzschild coordinates, i.e.,
\begin{align}\label{eq: Schwarzschild metric}
    g^\prime_{\mu\nu} = \mathrm{diag}\left(-\left(1-\frac{R_S}{r}\right),\left(1-\frac{R_S}{r}\right)^{-1},r^2,r^2 \sin^2\theta\right),
\end{align}
where \mbox{$R_S=2GM_E/c^2$} is the Schwarzschild radius and $M_E$ the Earth's mass.   We choose a static observer on (without loss of generality) the $x$-axis with a distance $R_E$, corresponding to the radius of the Earth, to the center of the mass distribution. The worldline in \textit{a priori coordinates}  $\mathsf{y}=(ct, r, \theta,\phi)$ with  coordinate time $t$ is then given by
\begin{align}\label{eq: worldline observer}
    \sigma^\alpha(t) = \left(c t, R_E, \frac{\pi}{2}, 0\right).
\end{align}
With that at hand one can easily derive the relation
\begin{align}
    \tau(t) = \sqrt{1-\epsilon}\;t,
\end{align}
where we have defined $\epsilon=R_S/R_E\approx 1.4\times 10^{-9}$. Later on, we will expand to first order in this small parameter, i.e. $\mathcal{O}(\epsilon)$.
Now, the orthonormal tetrad at $\tau=0$ can be constructed using \mbox{$e^\alpha_{(0)}=u^\alpha(0)$} and the orthonormality condition, Eq.~\eqref{eq: orthonormality condition}, yielding
\begin{subequations}
\begin{align}
    e^\alpha_{(0)}(0) &= \left(\sqrt{1-\epsilon}^{-1}, 0, 0, 0\right),\\
    e^\alpha_{(1)}(0) &= \left(0,\sqrt{1-\epsilon}, 0, 0\right),\\
    e^\alpha_{(2)}(0) &= \left(0, 0, R_E^{-1}, 0\right),\\
    e^\alpha_{(3)}(0) &= \left(0, 0, 0, R_E^{-1}\right).
\end{align}
\end{subequations}
Transporting this tetrad along $\sigma(\tau)$ via the transport equation, Eq.~\eqref{eq: Fermi Walker transport}, leads us to
\begin{subequations}
\begin{align}
    e^\alpha_{(\beta)}(\tau) &= e^\alpha_{(\beta)}(0).
\end{align}
\end{subequations}
Finally, we can calculate the metric in FWC using the Jacobian on the worldline, Eq.~\eqref{eq: Jacobian on worldline}, to express the four-acceleration $a^\mu(\tau)$ and the Riemannian curvature tensor $R_{\alpha\beta\gamma\delta}(\tau)$ in the Fermi-Walker basis. The metric in FWC up to second order is then given by
\begin{subequations}\label{eq: Metric in FWC}
    \begin{align}
    \begin{split}
        g_{00}(x^\alpha) &= -1-\left(\frac{x/R_E}{\sqrt{1-\epsilon}}-\frac{x^2/R_E^2}{1-\epsilon}\left(1-\frac{5\epsilon}{4}\right)+\frac{y^2+z^2}{2R_E^2}\right)\epsilon
    \end{split},\\
        g_{0i}(x^\alpha) &= 0,\\
        g_{11}(x^\alpha) &= 1+\frac{y^2+z^2}{6R_E^2}\epsilon,\\
        g_{22}(x^\alpha) &= 1+\frac{x^2-2z^2}{6R_E^2}\epsilon,\\
        g_{33}(x^\alpha) &= 1+\frac{x^2-2y^2}{6R_E^2}\epsilon,\\
        g_{ij}(x^\alpha) &= -\left(1-\frac{\delta^{i1}+\delta^{j1}}{2}\right) \frac{x^i x^j}{3R_E^2}\epsilon\;\textsf{for}\;i\neq j,
    \end{align}
\end{subequations}
where we used Eq.~\eqref{eq: Metric in FWC (general)} and denoted the FWC by $\mathsf{x}=(c\tau, x, y, z)$.

\section{Generators of Symmetry Group}
\label{Sec: Generators of Symmetry Group}
To determine the relativistic c.m. and relative coordinates for an atom made of constituent particles in curved spacetime, we need to find the single particle generators corresponding to the spacetime symmetry group attached to the oberserver's trajectory in the FWC~\cite{Osborn1968,Close1970,Liou1974,Krajcik1974}.
In flat spacetime for inertial motion of the observer, this corresponds then to the  Poincaré group~\cite{Dirac1949}. By finding the irreducible representation and the Casimir operators of the symmetry group, one obtains a global definition of a particle. The first Casimir operator $P^2$, i.e., the square of the four-momentum, defines the mass, and the second, i.e., the square of the Pauli-Lubanski pseudovector $W^2$, defines the spin of the particle~\cite{Wigner1939,Bargmann1948,Braathen1969}. For a system of non-interacting particles, the individual Poincaré generators are added independently, corresponding to the respective Hilbert space~\cite{Dirac1949,Macfarlane1963}. When adding interactions, the Casimir operators still exist but can become more complex as interactions can affect the symmetry properties and the definitions of conserved quantities.

When allowing for non-inertial motion, the global notion of particles does no longer hold as can be seen, for instance, by the Unruh effect~\cite{Unruh1976}. For an accurate description of spin in non-inertial frames, corrections that account for the non-inertial nature of the frame need to be included. As a consequence, the Casimir operator associated with the spin in local coordinates is no longer invariant under Lorentz transformations~\cite{Singh2007}. In fact, the spin depends on the corresponding Lorentz frame even for free particles~\cite{Peres2002}. Since the spin of a particle is defined locally with respect to a local inertial frame, it changes under local Lorentz transformations~\cite{Terashima2004}. The Casimir operator defining spin, i.e. $W^2$, possesses gravitational and frame dependent contributions~\cite{Singh2009,Singh2010}.
A possible approach is the use of local Lorentz transformations to investigate the spin properties of (possibly interacting) spin-$1/2$ particles in curved spacetime~\cite{Terashima2004,Alsing2009,Basso2021,Singh2007,Singh2009,Singh2010}.
Here however we will perform a perturbative ansatz to modify the known generators of the Poincaré group.

In this section, we determine the local Killing symmetries and map them to a perturbed Poincaré algebra to order $\mathcal{O}(\epsilon)$. This leads us to a local definition for the generators and the Casmir operators, cf., Refs.~\cite{Terashima2004,Alsing2009,Basso2021,Singh2007,Singh2009,Singh2010}. Consequently, there is only a local definition of a particle.

The generators of the spacetime symmetry group are the Killing vectors $\xi^\mu\partial_\mu$ whose components satisfy the Killing equations
\begin{align}\label{eq: Killing equation}
        \xi_{\beta;\alpha}+\xi_{\alpha;\beta}=0\Leftrightarrow\partial_\beta\xi_\alpha+\partial_\alpha\xi_\beta-2\Gamma^\gamma_{\alpha\beta}\xi_\gamma=0.
\end{align}
In flat spacetime the Killing equations are easily solved. Since the Christoffel symbols vanish in this case, the Killing equations reduce to
\begin{align}\label{eq: Killing equation flat spacetime}
        \partial_\beta\xi_\alpha+\partial_\alpha\xi_\beta=0.
\end{align}
The Killing vectors must then have the form
\begin{align}
        \xi_\alpha = a_\alpha + b_{\alpha\beta}x^\beta 
\end{align}
with $b_{\alpha\beta}=-b_{\beta\alpha}$ while $a_\alpha$ can be chosen arbitrarily. Therefore, there are ten linearly independent solutions: four translations in space and time through $a_\alpha$ and three rotations and boosts through $b_{\alpha\beta}$.

Solving the Killing equations exactly for curved spacetimes is in general very cumbersome.  We derive now GR corrections to the special relativistic generators for the metric derived in the previous section, Eq.~\eqref{eq: Metric in FWC}, by inserting the ansatz
\begin{align}\label{eq: Ansatz Killing equation}
    \xi_\alpha = \xi_{\textsf{SRT},\alpha} + \xi_{\textsf{GR},\alpha}\epsilon = a_\alpha + b_{\alpha\beta}x^\beta + c_{\alpha\beta\gamma}x^\beta x^\gamma + \mathcal{O}(x^3)
\end{align}
into the Killing equations, Eq.~\eqref{eq: Killing equation}, where $a_\alpha = a_{\textsf{SRT},\alpha} + a_{\textsf{GR},\alpha}$, $b_{\alpha\beta} = b_{\textsf{SRT},\alpha\beta} + b_{\textsf{GR},\alpha\beta}$ and $c_{\alpha\beta\gamma} = c_{\textsf{GR},\alpha\beta\gamma}$. For the case of observers in more general spacetimes, semi-analytical results may follow analogously. We can then insert the flat spacetime solutions for $a_{\textsf{SRT},\alpha}$ and $b_{\textsf{SRT},\alpha\beta}$, respectively, and find the corresponding GR correction terms $a_{\textsf{GR},\alpha}$, $b_{\textsf{GR},\alpha\beta}$ and $c_{\textsf{GR},\alpha\beta\gamma}$ of order $\mathcal{O}(\epsilon)$ (see Table~\ref{Table: Solution Killing vectors}) for the four translations in time and space
    \begin{align}\label{eq: Generator time translations}
    \begin{split}
    &H_\textsf{tot} = H_\textsf{SRT} + \epsilon \left(\frac{c^2 \tau}{R_E} p_x-\frac{1}{2}\left\{H_\textsf{SRT},\frac{x}{R_E}\right\}_+\right)
    \end{split},\\
    \begin{split}\label{eq: Generator space translations}
    &p_{\textsf{tot},k} = \frac{1}{2}\delta^{ij}\left\{p_i, g_{jk}\right\}_+ + \frac{\epsilon}{2}\left[\delta_{xk}\frac{\tau\;H_\textsf{SRT}}{R_E}\right. \\
    &\left. - \left(-2\right)^{\delta_{xk}}\left(\frac{c^2 \tau^2}{2R_E^2} p_k - \frac{1}{2}\left\{\frac{\tau\;H_\textsf{SRT}}{R_E},\frac{x^k}{R_E}\right\}_+\right)\right],
    \end{split}
    \end{align}
where $H_\textsf{SRT} = \sqrt{(mc^2)^2+(c\vec{p})^2}$ is the special-relativistic energy, and for the three rotations
    \begin{align}\label{eq: Rotation generator}
        \begin{split}
        J_{\textsf{tot},k} =&  L_{\textsf{SRT},k} + s_k \\
        &+\left(1-\delta_{xk}\right) \frac{\epsilon}{2}\left[\vec{e}_x\times\left(\frac{c^2 \tau^2}{2R_E} \vec{p} - \frac{1}{2}\left\{\frac{\tau\;H_\textsf{SRT}}{R_E},\vec{x}\right\}_+\right)\right]_k
        \end{split}
    \end{align}
where $\vec{L}_\textsf{SRT} = \vec{x}\times\vec{p}$ is the angular momentum in flat spacetime and boosts
    \begin{align}\label{eq: Boost generator}
    \begin{split}
        K_{\textsf{tot},k} &= \tau p_k - \frac{1}{2c^2}\left\{H_\textsf{SRT},x^k\left(1-\frac{\epsilon}{2}\frac{x}{R_E}\right)\right\}_+ - \frac{\left[\vec{p}\times\vec{s}\right]_k}{m c^2 + H_\textsf{SRT}}\\
        &-\frac{\epsilon}{2}\left[\left(1-\delta_{xk}\right)\frac{\tau}{R_E}\left[\vec{e}_x\times\vec{L}_{\textsf{SRT}}\right]_k - \delta_{xk} \frac{\tau^2}{2R_E}  H_\textsf{SRT}\right]
    \end{split}
    \end{align}
to first order in $\epsilon$, corresponding to the FWC frame attached to the observer,  cf. Eq.~\eqref{eq: worldline observer}, in Schwarzschild spacetime, Eq.~\eqref{eq: Metric in FWC}.
Since these single particle symmetry group generators are directly related to the corresponding metric, Eq.~\eqref{eq: Metric in FWC}, which itself depends on the observer's worldline, the gravitational corrections are observer-dependent as well.
Note that we need to include the usual spin  contributions in the  boost~\eqref{eq: Boost generator} and rotation generators~\eqref{eq: Rotation generator}, see e.g. Refs.~\cite{Osborn1968,Close1970,Liou1974,Krajcik1974}. If we were to neglect spin, the system of equations in order to determine the c.m. and relative coordinates would be overdetermined. However, since we are ultimately interested in systems without spin, we set the single particle spins to zero after solving them. Thus, we do not need to include gravitational corrections to the spin terms included in the symmetry generators. Intuitively, one would expect the replacement $\vec{p}\rightarrow\vec{p}_\textsf{tot}$ and $H_\textsf{SRT}\rightarrow H_\textsf{tot}$ in these spin terms, e.g.,
\begin{align}
    - \frac{\left[\vec{p}_\textsf{tot}\times\vec{s}\right]_k}{m c^2 + H_\textsf{tot}}
\end{align}
as spin term in Eq.~\eqref{eq: Boost generator}.  However, verifying this is beyond the scope of this paper and is left open for future work.

\begin{table}
	\caption{Components of the ten linearly independent Killing vectors \mbox{$\xi_\alpha = \xi_{\textsf{SRT},\alpha}+\xi_{\textsf{GR},\alpha}\epsilon$} and the corresponding symmetry \mbox{generator $G_i$}.}\label{Table: Solution Killing vectors}
	\centering
	\renewcommand{\arraystretch}{2}
	\begin{tabular}{l|c|c}
	$G_i$  &  $\xi_{\textsf{SRT},\alpha}$  &   $\xi_{\textsf{GR},\alpha}$   \\
    \hline\hline
    $H_\textsf{tot}$    &   $\left(1,0,0,0\right)$  &   $\left(\frac{-2x+y^2+z^2}{2R_E^2},\frac{c\tau}{R_E},0,0\right)$   \\
    \hline
    $p_{\textsf{tot},x}$    &   $\left(0,1,0,0\right)$    &   $\left(\frac{c\tau(R_E-2x)}{2R_E^2},\frac{3c^2\tau^2+2(y^2+z^2)}{6R_E^2},-\frac{xy}{3R_E^2},-\frac{xz}{3R_E^2}\right)$   \\
    \hline
    $p_{\textsf{tot},y}$    &   $\left(0,0,1,0\right)$    &   $\left(\frac{c\tau y}{2R_E^2},-\frac{xy}{3R_E^2},-\frac{3c^2\tau^2-4x^2+8z^2}{12R_E^2},\frac{2yz}{3R_E^2}\right)$\\
    \hline
     $p_{\textsf{tot},z}$   &   $\left(0,0,0,1\right)$  &   $\left(\frac{c\tau z}{2R_E^2},-\frac{xz}{3R_E^2},\frac{2yz}{3R_E^2},-\frac{3c^2\tau^2-4x^2+8y^2}{12R_E^2}\right)$\\
    \hline
     $J_{\textsf{tot},x}$   &   $\left(0,0,-z,y\right)$ &   $\left(0,0,0,0\right)$ \\
    \hline
     $J_{\textsf{tot},y}$   &   $\left(0,z,0,-x\right)$ &   $\left(\frac{c\tau z}{2R_E},0,0,-\frac{c^2\tau^2}{4R_E}\right)$ \\
    \hline
    $J_{\textsf{tot},z}$    &   $\left(0,-y,x,0\right)$ &   $\left(-\frac{c\tau y}{2R_E},0,\frac{c^2\tau^2}{4R_E},0\right)$ \\
    \hline
    $K_{\textsf{tot},x}$    &   $\left(-x,c\tau,0,0\right)$ &   $\left(\frac{c^2\tau^2-2x^2}{4R_E},0,0,0\right)$  \\
    \hline
    $K_{\textsf{tot},y}$    &   $\left(-y,0,c\tau,0\right)$    &   $\left(-\frac{xy}{2R_E},-\frac{c\tau y}{2R_E},\frac{c\tau x}{2R_E},0\right)$ \\
    \hline
    $K_{\textsf{tot},z}$    &   $\left(-z,0,0,c\tau\right)$ &   $\left(-\frac{xz}{2R_E},-\frac{c\tau z}{2R_E},0,\frac{c\tau x}{2R_E}\right)$
	\end{tabular}
	\renewcommand{\arraystretch}{1}
\end{table}

The Lie algebra for the generators $G_i$ of this modified Poincaré group, Eqs.~\eqref{eq: Generator time translations}--\eqref{eq: Boost generator}, can be written down in the form
\begin{align}\label{eq: Deformed Lie algebra}
    \left[G_i,G_j\right] = C_{ij}^k G_k + \frac{\epsilon}{2} \left(A_{ij}^k(x^\mu) G_k + G_k B_{ij}^k(x^\mu)\right),
\end{align}
where $C_{ij}^k$ are the well-known Poincaré structure constants. Eq.~\eqref{eq: Deformed Lie algebra} describes a deformation of the Poincaré algebra to first order in $\epsilon$. Deformed Lie algebras~\cite{Levi-Nahas1967,Lukierski1992,Bacry1993,Maslanka1993,Fernandez1996,Zhang2012,Pfeifer2022} have already been investigated in the context of quantum gravity, i.e., the attempt to unify quantum mechanics with GR. A common approach of this attempt is the non-commutativity of spacetime coordinates~\cite{Bacry1993,Pfeifer2022}, which leads to a deformation of the commutation relations between the generators and the Casimir operators of the symmetry group, in order to accommodate the Poincaré and Heisenberg algebra. 

\section{Generalized c.m. and Relative Coordinates}
\label{Sec: Generalized c.m. and relative Coordinates}
In this section we use the single particle coordinate frame defined in Sec.~\ref{Sec: Metric in FWC}, i.e., the Fermi-Walker frame, and the single particle symmetry group generators found in Sec.~\ref{Sec: Generators of Symmetry Group} to derive general-relativistic correction terms for the c.m. and relative coordinates of an atom consisting of two particles.
\subsection{Calculation of relativistic c.m. coordinates}
We can now proceed to calculate gravitational corrections to the relativistic c.m. and relative coordinates. The underlying principle is that the sum of the single particle symmetry generators $G_i$ has to have the same form in relativistic c.m. coordinates as the single particle generators~\cite{Osborn1968,Close1970,Liou1974,Krajcik1974}, i.e.,
\begin{align}
    \sum_{j}G_i(\vec{x}_j,\vec{p}_j,\vec{s}_j,m_j) = G_i(\vec{R},\vec{P},\vec{S},M)\quad\text{for}\quad i=1,...,10.
\end{align}
Although this works for an arbitrary number of particles, we restrict our calculations to hydrogenoid atoms, consisting of two particles with, e.g., an electron and a proton. In this way we arrive at ten equations that can be used to find the relativistic c.m. position $\vec{R}$, the total momentum $\vec{P}$, spin $\vec{S}$ and mass $M$. 

Then we can insert the ansatz 
\begin{subequations}
\begin{align}\label{eq:c.m. ansatz}
    \vec{R}&=\vec{R}_{\textsf{NR}}+\vec{R}_{\textsf{SRT}}+\vec{R}_\textsf{GR}\; \epsilon,\\
    \vec{P}&=\vec{P}_{\textsf{NR}}+\vec{P}_{\textsf{SRT}}+\vec{P}_\textsf{GR}\; \epsilon,\\
    \vec{S}&=\vec{S}_{\textsf{NR}}+\vec{S}_{\textsf{SRT}}+\vec{S}_\textsf{GR}\; \epsilon,\\
    M&=M_{\textsf{NR}}+M_{\textsf{SRT}}+M_\textsf{GR}\; \epsilon,
\end{align}
\end{subequations}
where the subscripts $\textsf{NR}$, $\textsf{SRT}$ and $\textsf{GR}$ labels the non-relativistic (NR) solutions and the special (SRT) and GR corrections, respectively.
Expanding all ten equations in $\epsilon$, we see that the zeroth order is already solved by the NR and the SRT solutions that are already known~\cite{Osborn1968,Close1970,Liou1974,Krajcik1974}. We can then use the first order equations to calculate the gravitational correction terms. Since we are only interested in terms up to the order $c^{-2}$ and the Schwarzschild radius $R_S$ is already of this order, we only need to solve the equations up to the zeroth order in $c^{-1}$. Thus, we arrive at the correction for the c.m. position
\begin{align}\label{eq:c.m. GR correction}
    \vec{R}_\textsf{GR} = -\frac{\mu}{2M}\frac{r_{\textsf{NR},x}}{R_E} \vec{r}_\textsf{NR} - \frac{\tau}{2MR_E}\vec{L}_\textsf{int,NR}\times\vec{e}_x,
\end{align}
where $\vec{r}_\textsf{NR}=\vec{r}_1-\vec{r}_2$ is the non-relativistic relative coordinate, $\vec{L}_\textsf{int,NR}=\vec{r}_\textsf{NR}\times\vec{p}_\textsf{NR}$ is the non-relativistic internal angular momentum and $M=m_1+m_2$ and $\mu=m_1m_2/M$ are the total and the reduced mass of the two-particle system, respectively. The total momentum $\vec{P}$ and the mass do not acquire any gravitational corrections to our order of approximation. The spin $\vec{S}$ does get gravitational corrections. However, since we are not interested in spin dynamics, we can ignore them.

\subsection{Calculation of relativistic relative coordinates}
With the relativistic c.m. coordinates we can compute the relativistic relative coordinates. We assume that the relativistic c.m. coordinates are connected to the non-relativistic coordinates via a unitary transformation~\cite{Liou1974}
\begin{subequations}
\begin{align}
    \vec{P}&=\ee^{\frac{\ii}{\hbar} u}\vec{P}_\textsf{NR}\ee^{-\frac{\ii}{\hbar} u},\\
    \vec{R}&=\ee^{\frac{\ii}{\hbar} u}\vec{R}_\textsf{NR}\ee^{-\frac{\ii}{\hbar} u},\\
    \vec{r}&=\ee^{\frac{\ii}{\hbar} u}\vec{r}_\textsf{NR}\ee^{-\frac{\ii}{\hbar} u},\\
    \vec{p}&=\ee^{\frac{\ii}{\hbar} u}\vec{p}_\textsf{NR}\ee^{-\frac{\ii}{\hbar} u},
\end{align}
\end{subequations}
where we know that $\vec{P}=\vec{P}_\textsf{NR}=\vec{p}_1+\vec{p}_2$, so that the generator of the unitary transformation $u$ is no function of the non-relativistic c.m. position $\vec{R}_\textsf{NR}$. However, we already know $\vec{R}$ through Eqs.~\eqref{eq:c.m. ansatz} and~\eqref{eq:c.m. GR correction} up to the order $c^{-2}$. We can thus expand $u$ and $\vec{R}$ in orders of $c^{-2}$, i.e.,
\begin{subequations}
\begin{align}
    \vec{R}&=\vec{R}_\textsf{NR}+\vec{R}_2+\mathcal{O}(c^{-4}),\\
    u&=u_2+\mathcal{O}(c^{-4}),
\end{align}
\end{subequations}
and use that $\vec{R}_\textsf{NR}=\ii\hbar\frac{\partial}{\partial\vec{P}}$ in momentum representation to arrive at
\begin{align}
   \frac{\partial u_2}{\partial\vec{P}}=\vec{R}_2,
\end{align}
which leads us directly to the second order solution
\begin{align}\label{eq: unitary transformation relative coordinates}
\begin{split}
    &u_2 = c^{-2}\left[-\frac{\Delta m}{4\mu M^2}\left(\vec{p}_\textsf{NR}^2\left(\vec{r}_\textsf{NR}\cdot\vec{P}\right)+\textsf{h.c.}\right)\right.\\
        &\left.+\frac{\left(\vec{P}\cdot\vec{p}_\textsf{NR}\right)\left(\vec{P}\cdot\vec{r}_\textsf{NR}\right)+\textsf{h.c.}}{4M^2}\right]\\
        &-\underbrace{\epsilon\left[\frac{\mu}{2M}\frac{r_{\textsf{NR},x}}{R_E}\left(\vec{P}\cdot\vec{r}_\textsf{NR}\right)+\frac{\tau}{2MR_E}\left(\vec{P}\cdot\left(\vec{L}_\textsf{int,NR}\times\vec{e}_x\right)\right)\right]}_{\textsf{GR corrections}},
\end{split}
\end{align}
where we introduced the mass difference $\Delta m = m_1-m_2$. We now have determined $u$ up to the order $c^{-2}$ and can use it to calculate the relativistic relative coordinates
\begin{subequations}
    \begin{align}
        \vec{r} &= \vec{r}_\textsf{NR} + \vec{r}_\textsf{SRT} + \vec{r}_\textsf{GR}\; \epsilon,\\
        \vec{p} &= \vec{p}_\textsf{NR} + \vec{p}_\textsf{SRT} + \vec{p}_\textsf{GR}\; \epsilon
    \end{align}
\end{subequations}
via Baker-Campbell-Hausdorff formula, which finally leads us to the gravitational corrections for the relative coordinates
\begin{subequations}\label{eq: relativistic relative Coordinates}
\begin{align}
    \vec{r}_\textsf{GR} =&  \frac{\tau}{2MR_E}\left[\left(\vec{P}\cdot\vec{r}_\textsf{NR}\right)\vec{e}_x - r_{\textsf{NR},x} \vec{P}\right],\\
    \begin{split}
    \vec{p}_\textsf{GR} =& \frac{\mu}{2MR_E}\left[r_{\textsf{NR},x}\vec{P}+\left(\vec{P}\cdot\vec{r}_\textsf{NR}\right)\vec{e}_x\right]\\
                    & + \frac{\tau}{2MR_E}\left[\left(\vec{P}\cdot\vec{p}_\textsf{NR}\right)\vec{e}_x - p_{\textsf{NR},x} \vec{P}\right].
    \end{split}
\end{align}
\end{subequations}

\section{Two-Particle Dynamics in Curved Spacetime}
\label{Sec: Two-Particle Dynamics in Curved Spacetime}
Having determined the gravitational corrections to relative and c.m. coordinates, in the next step we want to study the dynamics of a localized matter system in the presence of gravity and light. This is in contrast to the usual approaches, for instance when investigating atom interferometry, where a plethora of gravitationally induced correction terms arise and need to be sorted.

As a simple example, we consider in the following a hydrogenoid two-particle atom, consisting of, e.g., an electron and a proton, coupled to internal and external electromagnetic fields and gravitation.
The calculations of the following Sections~\ref{sec: Total Lagrangian} and~\ref{sec: Total Hamiltonian} as well as the Appendix~\ref{App: Maxwell equations} are in total analogy to the calculations done by Schwartz in Refs.~\cite{Schwartz2019,Schwartz2020b}. Here however, we consider a Schwarzschild observer on the surface of the Earth, Eq.~\eqref{eq: Metric in FWC}, in contrast to being located at infinite distance, and determine the Hamiltonian in the gravitationally corrected c.m. and relative coordinates, Eqs.~\eqref{eq:c.m. GR correction} and~\eqref{eq: relativistic relative Coordinates}.
\subsection{Total Lagrangian}\label{sec: Total Lagrangian}

Starting with the Lagrangian, we may write
\begin{align}
    L = L_\textsf{kin} + L_\textsf{em}
\end{align}
for the kinetic part, and the coupling of the atom to internal and external electromagnetic fields including gravitational effects.
The kinetic Lagrangian $L_\textsf{kin}$ can be found in a straight forward manner by adding the classical kinetic Lagrangian for single point particles with masses $m_i$ and positions $\vec{r}_i$, i.e.,
\begin{align}
\begin{split}
    L_\textsf{kin} &=\!\sum_{i=1}^2\!\left(-m_ic^2\sqrt{-g_{\mu\nu}\dot{r}_i^\mu\dot{r}_i^\nu/c^2}\right)\\
    &=\!\sum_{i=1}^{2}\!\left[\!\frac{1}{2}m_i\dot{\vec{r}}_i^2\left(1+\frac{\dot{\vec{r}}_i^2}{4c^2}\right)-\frac{GM_Em_ix_i}{R_E^2}\left(1+\frac{\dot{\vec{r}}_i^2}{2c^2}\right)-m_ic^2\!\right]\!.
\end{split}
\end{align}
Note, that the index $i$ labels the respective particle in this context.

To calculate the electromagnetic Lagrangian
\begin{align}\label{eq: em. Lagrangian}
    L_\textsf{em} = \int\mathrm{d}^3r\left(-\frac{\epsilon_0 c^2}{4}\sqrt{-g}F_{\textsf{tot}\,\mu\nu}F_{\textsf{tot}}^{\mu\nu} + j^\mu A_{\textsf{tot}\,\mu}\right),
\end{align}
including the interaction of light and matter we firstly have to couple the electromagnetic to the gravitational field, where $g$ is the determinant of the metric $g_{\mu\nu}$, $j^\mu$ is the four-current density of the two-particle system, $A_{\textsf{tot}\,\mu}$ is the electromagnetic four-potential and $F_{\textsf{tot}\,\mu\nu}=\partial_\mu A_{\textsf{tot.}\nu}-\partial_\nu A_{\textsf{tot.}\mu}$ is the electromagnetic field tensor. For the derivation of this electromagnetic Lagrangian we need to solve Maxwell's equations in curved spacetime for the internal as well as the external electromagnetic fields. This is done in Appendix~\ref{App: Maxwell equations}. We can split the electromagnetic field tensor $F_{\textsf{tot}\,\mu\nu}$ as well as the electromagnetic four-potential $A_{\textsf{tot}\,\mu}$ into an internal and an external part, i.e.,
\begin{subequations}
\begin{align}
    F_{\textsf{tot}\,\mu\nu} &= \mathcal{F}_{\mu\nu} + F_{\mu\nu},\\
    A_{\textsf{tot}\,\mu}    &= \mathcal{A}_\mu      + A_\mu
\end{align}
\end{subequations}
and plug in the solutions found in Appendix~\ref{App: Maxwell equations}. Expanding all to the appropriate order of $c^{-2}$ leads us to
\begin{widetext}
\begin{align}
    \begin{split}
        L_\textsf{em,ext} =& -\frac{\epsilon_0}{2}\int\mathrm{d}^3r \sqrt{-g}\left[g^{00}\delta_{ab}(\partial_tA^{\perp a})(\partial_tA^{\perp b})+c^2(\vec{\nabla}\times \vec{A}^\perp)^2\right]\\
        = &\frac{\epsilon_0}{2}\int\mathrm{d}^3r\left[\left(1-\frac{GM_E}{R_E^2c^2}x\right)(\partial_t\vec{A}^\perp)^2-\left(1+\frac{GM_E}{R_E^2c^2}x\right)c^2(\vec{\nabla}\times \vec{A}^\perp)^2\right]
    \end{split}
\end{align}
for the purely external part of the Lagrangian and
\begin{align}
        \begin{split}
            L_\textsf{em,ex-int} &= \int\mathrm{d}^3r j^\mu A_\mu^\perp - \frac{\epsilon_0c^2}{2}\int\mathrm{d}^3r\sqrt{-g}\mathcal{F}_{\mu\nu}F^{\mu\nu}\\
            &= \int\mathrm{d}^3r \vec{j}\cdot\vec{A}^\perp + \epsilon_0 \int\mathrm{d}^3r \phi_\textsf{el}(\partial_t A^{\perp a})(\partial_a\sqrt{-g}g^{00}) + \epsilon_0\int\mathrm{d}^3r \left[(\partial_t\vec{\mathcal{A}}^\perp)(\partial_t\vec{A}^\perp)-c^2(\vec{\nabla}\times\vec{\mathcal{A}}^\perp)(\vec{\nabla}\times\vec{A}^\perp)\right]
        \end{split}
\end{align}
for the extern-intern cross terms of the Lagrangian, where the last summand can be neglected since it is a diverging back reaction term (cf. Sonnleitner et al.~\cite{Sonnleitner2018}) to arrive at
\begin{align}
    L_\textsf{em,ex-int} = \int\mathrm{d}^3r \vec{j}\cdot\vec{A}^\perp + \epsilon_0 \int\mathrm{d}^3r \frac{G M_E}{R_E^2 c^2}\phi_\textsf{el}(\partial_t A^{\perp 1}).
\end{align}
The kinetic Maxwell term of the purely internal part of the electromagnetic Lagrangian
\begin{align}
    L_\textsf{em,int} = \int\mathrm{d}^3r j^\mu\mathcal{A}^\perp_\mu - \frac{\epsilon_0c^2}{4}\int\mathrm{d}^3r \sqrt{-g}\mathcal{F}_{\mu\nu}\mathcal{F}^{\mu\nu}
\end{align}
can be simplified via integration by parts, i.e.,
\begin{align}
    \begin{split}
        - \frac{\epsilon_0c^2}{4}\!\int\!\mathrm{d}^3r \sqrt{-g}\mathcal{F}_{\mu\nu}\mathcal{F}^{\mu\nu} &= -\frac{\epsilon_0c^2}{2}\!\int\!\mathrm{d}^3r \sqrt{-g}(\partial_\mu \mathcal{A}^\perp_\nu)\mathcal{F}^{\mu\nu} \\
        &=\frac{\epsilon_0c^2}{2}\left[\!\int\!\mathrm{d}^3r \mathcal{A}^\perp_\nu (\partial_a\sqrt{-g}\mathcal{F}^{a\nu})-\!\int\!\mathrm{d}^3r \sqrt{-g}(\partial_0\mathcal{A}^\perp_\nu)\mathcal{F}^{0\nu}\right]\\
        &= \frac{\epsilon_0c^2}{2}\!\int\!\mathrm{d}^3r \sqrt{-g}\mathcal{A}^\perp_\nu \nabla_\mu\mathcal{F}^{\mu\nu} - \frac{\epsilon_0c^2}{2}\!\int\!\mathrm{d}^3r (\partial_0 \sqrt{-g}\mathcal{A}^\perp_\nu\mathcal{F}^{0\nu}).
    \end{split}
\end{align}
We can then use Maxwell's equations for the first summand, whereas the second summand is of the order $\mathcal{O}(c^{-4})$, cf. Eq.~(4.4.37) in Ref.~\cite{Schwartz2020b}. Therefore, the purely internal electromagnetic Lagrangian reduces to
\begin{align}
\begin{split}
   L_\textsf{em,int} = \frac{1}{2}\int\mathrm{d}^3r j^\mu \mathcal{A}_\mu = \frac{1}{2}\int\mathrm{d}^3r\left(\vec{j}\cdot\vec{\mathcal{A}}^\perp - \frac{e_1 e_2}{2\pi\epsilon_0\vert\vec{r}_1-\vec{r_2}\vert}\left(1+\frac{G M_E}{2R_E^2 c^2}(x_1+x_2)\right)\right),
\end{split}
\end{align}
where $x_i$ is the $x$-component of $\vec{r}_i$ and the total electromagnetic Lagrangian reads
\begin{align}
    \begin{split}
           &L_\textsf{em} = L_\textsf{em,int} + L_\textsf{em,ext} + L_\textsf{em,ext-int}\\
           =&\frac{1}{2}\int\mathrm{d}^3r\left(\vec{j}\cdot\vec{\mathcal{A}}^\perp - \frac{e_1 e_2}{2\pi\epsilon_0\vert\vec{r}_1-\vec{r_2}\vert}\left(1+\frac{G M_E}{2R_E^2 c^2}(x_1+x_2)\right)\right) +\!\int\!\mathrm{d}^3r \vec{j}\cdot\vec{A}^\perp\\
           &+ \frac{\epsilon_0}{2}\!\int\!\mathrm{d}^3r\left[\left(1-\frac{GM_E}{R_E^2c^2}x\right)(\partial_t\vec{A}^\perp)^2-\left(1+\frac{GM_E}{R_E^2c^2}x\right)c^2(\vec{\nabla}\times \vec{A}^\perp)^2\right] + \epsilon_0 \int\mathrm{d}^3r \frac{G M_E}{R_E^2 c^2}\phi_\textsf{el}(\partial_t A^{\perp 1}) \\
           =&-\frac{e_1e_2}{4\pi\epsilon_0 \vert\vec{r}_1-\vec{r}_2\vert} \! \left(1-\frac{1}{2c^2}\!\left(\dot{\vec{r}}_1\cdot\dot{\vec{r}}_2+\frac{[\dot{\vec{r}}_1\cdot(\vec{r}_1-\vec{r}_2)][\dot{\vec{r}}_2\cdot(\vec{r}_1-\vec{r}_2)]}{\vert\vec{r}_1-\vec{r}_2\vert^2} -\frac{G M_E}{R_E^2}(x_1+x_2)\right)\!\right) + e_1 \dot{\vec{r}}_1\vec{A}^\perp(\vec{r}_1) + e_2 \dot{\vec{r}}_2\vec{A}^\perp(\vec{r}_2) \\
           &+ \frac{\epsilon_0}{2}\int\mathrm{d}^3r\left[\left(1-\frac{GM_E}{R_E^2c^2}x\right)(\partial_t\vec{A}^\perp)^2-\left(1+\frac{GM_E}{R_E^2c^2}x\right)c^2(\vec{\nabla}\times \vec{A}^\perp)^2\right] + \epsilon_0 \int\mathrm{d}^3r \frac{G M_E}{R_E^2 c^2}\phi_\textsf{el}(\partial_t A^{\perp 1}),
    \end{split}
\end{align}
where we used the definition of the current density, Eq.~\eqref{eq: current density}, and the internal electromagnetic fields, Eqs.~\eqref{eq: lowest order internal scalar potential} and~\eqref{eq: lowest order internal vector potential} in the last step.

The total Lagrangian including all couplings between the atoms, the electromagnetic and the gravitational field is then given by
\begin{align}\label{eq: total Lagrangian}
    \begin{split}
            &L = L_\textsf{kin} + L_\textsf{em}= \sum_{i=1}^{2}\left[\frac{1}{2}m_i\dot{\vec{r}}_i^2\left(1+\frac{\dot{\vec{r}}_i^2}{4c^2}\right)-\frac{GM_Em_ix_i}{R_E^2}\left(1+\frac{\dot{\vec{r}}_i^2}{2c^2}\right)-m_ic^2\right]\\
            &-\frac{e_1e_2}{4\pi\epsilon_0 \vert\vec{r}_1-\vec{r}_2\vert} \! \left(1-\frac{1}{2c^2}\!\left(\dot{\vec{r}}_1\cdot\dot{\vec{r}}_2+\frac{[\dot{\vec{r}}_1\cdot(\vec{r}_1-\vec{r}_2)][\dot{\vec{r}}_2\cdot(\vec{r}_1-\vec{r}_2)]}{\vert\vec{r}_1-\vec{r}_2\vert^2} -\frac{G M_E}{R_E^2}(x_1+x_2)\right)\!\right) + e_1 \dot{\vec{r}}_1\vec{A}^\perp(\vec{r}_1) + e_2 \dot{\vec{r}}_2\vec{A}^\perp(\vec{r}_2) \\
           &+ \frac{\epsilon_0}{2}\int\mathrm{d}^3r\left[\left(1-\frac{GM_E}{R_E^2c^2}x\right)(\partial_t\vec{A}^\perp)^2-\left(1+\frac{GM_E}{R_E^2c^2}x\right)c^2(\vec{\nabla}\times \vec{A}^\perp)^2\right] + \epsilon_0 \int\mathrm{d}^3r \frac{G M_E}{R_E^2 c^2}\phi_\textsf{el}(\partial_t A^{\perp 1}).
    \end{split}
\end{align}
\end{widetext}
\subsection{Total Hamiltonian}\label{sec: Total Hamiltonian}
We can now perform the Legendre transformation of the total Lagrangian, Eq.~\eqref{eq: total Lagrangian}, to arrive at the Hamiltonian
\begin{align}
    H = \sum_{i=1}^2 \vec{p}_i\cdot\dot{\vec{r}}_i + \vec{\Pi}^\perp\cdot(\partial_t \vec{A}^\perp) - L,
\end{align}
with the canonical momenta
\begin{align}
    \vec{p}_i &= \frac{\partial L}{\partial \dot{\vec{r}}_i},\\
    \vec{\Pi}^\perp &= \frac{\delta L}{\delta (\partial_t \vec{A}^\perp)}.
\end{align}
Using the generalized relativistic c.m. and relative coordinates, Eqs.~\eqref{eq:c.m. GR correction} and~\eqref{eq: relativistic relative Coordinates}, and performing the Power-Zienau-Woolley transformation, cf. Refs.~\cite{Sonnleitner2018,Andrews2018,Woolley2020,Schwartz2019,Schwartz2020b}, i.e., the replacement
\begin{subequations}
    \begin{align}
        \vec{p}_i - e_i\vec{A}^\perp(\vec{r}_i) &\rightarrow \vec{p}_i + \frac{\vec{d}\times\vec{B}^\perp(\vec{R})}{2},\\
        \vec{\Pi}^\perp(\vec{r}) &\rightarrow\tilde{\vec{\Pi}}^\perp(\vec{r}) + \vec{\mathcal{P}}^\perp(\vec{r}),
    \end{align}
\end{subequations}
where $\vec{B}^\perp = \vec{\nabla}\times\vec{A}^\perp$ is the magnetic field, 
\begin{align}\label{eq: polarization field}
    \begin{split}
    \vec{\mathcal{P}}(\vec{r},t) = \sum_{i=1}^2 e_i &[\vec{r}_i(t) - \vec{R}(t)]\times\\
    &\times\int_0^1\mathrm{d}\lambda\, \delta\big(\vec{r}-\vec{R}(t)-\lambda[\vec{r}_i(t)-\vec{R}(t)]\big)    
    \end{split}
\end{align}
is the polarization field and $\vec{d} = \sum_{i=1}^{2}e_i\vec{r}_i$ is the electric dipole moment. By means of the generalized relativistic c.m. and relative coordinates of the previous section, and after the dipole approximation, we arrive at the total Hamiltonian
\begin{align}\label{eq: total Hamiltonian}
    H = H_\textsf{c.m.} + H_\textsf{int} + H_\textsf{AL} + H_\textsf{L} + H_\textsf{X}
\end{align}
\begin{widetext}
with
\begin{align}
\begin{split}
    H_\textsf{c.m.} = \frac{\vec{P}^2}{2M}\left[1-\frac{1}{Mc^2}\left(\frac{\vec{p}^2}{2\mu}-\frac{e^2}{4\pi\epsilon_0r}\right)\right] + M \left[1+\frac{1}{Mc^2}\left(\frac{\vec{p}^2}{2\mu}-\frac{e^2}{4\pi\epsilon_0r}\right)\right]\phi(\vec{R}) - \frac{\vec{P}^4}{8M^3c^2} + \frac{1}{2Mc^2}\vec{P}\cdot\phi(\vec{R})\vec{P}
\end{split}
\end{align}
being the c.m. Hamiltonian including the mass defect $M\rightarrow M+H_\textsf{int}/c^2$, where
\begin{align}\label{eq: internal Hamiltonian}
\begin{split}
    H_\textsf{int} = \frac{\vec{p}^2}{2\mu} - \frac{e^2}{4\pi\epsilon_0r} - \frac{m_1^3+m_2^3}{M^3}\frac{\vec{p}^4}{8\mu^3c^2} - \frac{e^2}{4\pi\epsilon_0}\frac{1}{2\mu M^2c^2}\left[\vec{p}\frac{1}{r}\vec{p} + (\vec{p}\cdot\vec{r})\frac{1}{r^3}(\vec{r}\cdot\vec{p})\right] 
\end{split}
\end{align}
is the atomic internal Hamiltonian expanded around the observer's position, Eq.~\eqref{eq: worldline observer}, to second order in FWC. The equation
\begin{align}\label{eq: Atom-light Hamiltonian}
\begin{split}
    H_\textsf{AL} &= \left(1 + \frac{\phi(\vec{R})}{c^2}\right)\frac{\tilde{\vec{\Pi}}^\perp}{\epsilon_0}\cdot \vec{d} + \frac{1}{2M}\left[\vec{P}\cdot(\vec{d}\times\vec{B}^\perp(\vec{R}))+\textsf{h.c.}\right]-\frac{m_1-m_2}{4\mu M}\left[\vec{p}\cdot(\vec{d}\times\vec{B}^\perp(\vec{R}))+\textsf{h.c.}\right] + \frac{1}{8\mu}\left(\vec{d}\times\vec{B}^\perp(\vec{R})\right)^2\\
    &+ \frac{1}{2\epsilon_0}\int\mathrm{d}^3r \left(1 + \frac{\phi(\vec{r})}{c^2}\right) \vec{\mathcal{P}}^{\perp 2}_d(\vec{r},t) - \int\mathrm{d}^3r \phi_\textsf{el}(\vec{r})\frac{\vec{\nabla}\phi(\vec{r})}{c^2}\left(\vec{\tilde{\Pi}}^\perp + \vec{\mathcal{P}}_d^\perp\right)
\end{split}
\end{align}
is the Hamiltonian describing atom-light interaction including gravitational corrections, where $\vec{\mathcal{P}}^{\perp}_d(\vec{r},t) = \vec{d}\delta(\vec{r}-\vec{R})$ is the dipole approximated polarization field, Eq.~\eqref{eq: polarization field}, for a vanishing total charge, i.e. $\sum_i e_i = 0$. The equation
\begin{align}
\begin{split}
    H_\textsf{L} &= \frac{\epsilon_0}{2}\int\mathrm{d}^3r \left(1 + \frac{\phi(\vec{r})}{c^2}\right)\left[\left(\frac{\Tilde{\vec{\Pi}}^\perp}{\epsilon_0}\right)^2+c^2\vec{B}^{\perp 2}\right]
\end{split}
\end{align}
is the electromagnetic field energy. Finally, we have
\begin{align}\label{eq: Cross-term Hamiltonian}
    H_\textsf{X} = - \frac{1}{4Mc^2}\left[(\vec{P}\cdot\vec{p})(\vec{r}\cdot\vec{\nabla}\phi(\vec{R}))+\textsf{h.c.}\right] - \frac{3}{4Mc^2}\left[(\vec{P}\cdot\vec{r})(\vec{p}\cdot\vec{\nabla}\phi(\vec{R}))+\textsf{h.c.}\right],
\end{align}
coupling the internal and external dynamics of the atom.
\end{widetext}

Comparing this Hamiltonian with the Hamiltonian of Schwartz and Giulini, Refs.~\cite{Schwartz2019,Schwartz2020b}, we see that -- by employing general-relativistically corrected c.m. and relative operators -- the cross-coupling between the internal dynamics and the gravitational field, i.e., the terms
\begin{align}
        -\frac{1}{2c^2}&\frac{m_1-m_2}{m_1m_2}\vec{p}\cdot\left(\vec{r}\cdot\vec{\nabla}\phi(\vec{R})\right)\vec{p}\\
        \frac{1}{c^2}&\frac{m_1-m_2}{M}\frac{e^2}{8\pi\epsilon_0 r}\vec{r}\cdot\vec{\nabla}\phi(\vec{R})
\end{align}
are vanishing. However, we find a differing cross-coupling Hamiltonian, Eq.~\eqref{eq: Cross-term Hamiltonian}. When applying non-relativistic c.m. and relative coordinates, we find the same result like in Refs.~\cite{Schwartz2019,Schwartz2020b} with $\beta=\gamma=0$. Thus, to our order of approximation, the internal energy levels of the atom do not shift due to gravity, c.f., Eq.~\eqref{eq: internal Hamiltonian}. When measuring the internal structure of an atom, we need to couple an electro-magnetic field to the atom, e.g., through a Rabi cycle~\cite{Rabi1937}. Since the prefactors of the atom-light interaction terms, Eq.~\eqref{eq: Atom-light Hamiltonian}, undergo gravitational corrections, one will observe a change of the transition rates. Importantly, one would only find a change of transition rates of the transitions that would also occur in the absence of gravity. There are no additional transitions induced purely by gravity. Furthermore, this change of the transition rates depends on the c.m. position of the atom. In the context of atom interferometry, this effect would lead in principle to measurable effects. Atoms located on different heights/interferometer arms will feel different Rabi frequencies.
As in the case of special-relativistic corrections, cf. Ref.~\cite{Sonnleitner2018}, the mass defect becomes implicit as can be seen via
\begin{align}\label{eq: c.m. Hamiltonian mass defect}
\begin{split}
    H_\textsf{c.m.} =& \frac{\vec{P}^2}{2M} \left[1 - \frac{H_\textsf{int}}{Mc^2}\right] + M \left[1 + \frac{H_\textsf{int}}{Mc^2}\right] \phi(\vec{R}) - \frac{\vec{P}^4}{8M^3c^2}\\
    &+ \frac{1}{2Mc^2}\vec{P} \cdot \phi(\vec{R}) \vec{P}.
\end{split}
\end{align}
The mass defect plays a key role in atom interferometric tests of the universality of free fall and the universality of gravitational redshift, c.f., Refs.~\cite{Loriani2019,Roura2020,Ufrecht2020,DiPumpo2021,DiPumpo2022,DiPumpo2023,Janson2024}.
Thus, this work provides the basis for a theoretical description of general relativistic effects based on first principles and relates our findings with actual laboratory quantities by considering a reference frame of a local observer on the surface of the Earth.
Note, that we do not observe quadratic terms in the (shifted) gravitational potential $\phi(\vec{R})=G M_E R^x/R_E^2$, cf. Eq.~(5.5) in Ref.~\cite{Schwartz2019} and Eq.~(4.5.10b) in Ref.~\cite{Schwartz2020b} since we only expanded to first order in $\epsilon = R_S/R_E$. Moreover, in contrast to the internal Hamiltonian obtained by Schwartz and Giulini, Eq.~(5.6) in Ref.~\cite{Schwartz2019} and Eq.~(4.5.10c) in Ref.~\cite{Schwartz2020b}, the internal atomic Hamiltonian, Eq.~\eqref{eq: internal Hamiltonian}, does not contain gravitational correction terms.
Furthermore, when comparing the results of Schwartz and Giulini~\cite{Schwartz2019,Schwartz2020b} and our result, Eq.~\eqref{eq: internal Hamiltonian} with the Hamiltonian of Parker, Eq.~(9.13) in Ref.~\cite{Parker1980}, we see that neither the Hamiltonian of Schwartz and Giulini nor our Hamiltonian contains the term leading to internal geodesic deviation forces, i.e., (adapted to our nomenclature)
\begin{align}
    \frac{1}{2} \mu R_{0l0m} r^l r^m
\end{align}
generated by gravitational gradients. This is because we (as well as Schwartz and Giulini~\cite{Schwartz2019,Schwartz2020b}) expanded the gravitational field around the c.m. position only to the first order in relative coordinates. In fact, when expanding to higher orders, the effect of geodesic deviation forces would be present in the final Hamiltonian. Nevertheless, since we are dealing with the scenario of a weakly curved spacetime, the gravitational gradient is expected to be small. Therefore, and due to parity arguments, we can neglect this effect, when calculating internal energy shifts induced by gravity.

Even when choosing the general-relativistically corrected c.m./relative atomic operators, there remains a cross term coupling internal and external atomic degrees of freedom. To determine the full atomic dynamics, one can perform a unitary transformation reversing the gravitational part of the unitary transformation, Eq.~\eqref{eq: unitary transformation relative coordinates}, i.e.,
\begin{align}\label{eq: unitary transformation cross-term}
    H^\prime = \ee^{\frac{\ii}{\hbar} \tilde{u}} H \ee^{-\frac{\ii}{\hbar} \tilde{u}}\;\text{with}\,\tilde{u} = \epsilon\frac{\mu}{2M}\frac{r_{\textsf{NR},x}}{R_E}\left(\vec{P}\cdot\vec{r}_\textsf{NR}\right),
\end{align}
where it is necessary to assume that the transversal motion of the atom is frozen-in, i.e. $P_y$, $P_z=0$.
One can then compute the dynamics in that frame without the presence of the Hamiltonian $H_\textsf{X}$, and subsequently reverse the unitary. 
In fact, the transformed picture then corresponds to the one representing laboratory setups, such that experimental settings and initial states are prepared in that picture.
Thus, in quasi-1D settings, we do expect negligible effects originating from the cross-term Eq.~\eqref{eq: Cross-term Hamiltonian}, e.g., in the context of atom interferometry.

\section{Conclusion}
\label{Sec: Conclusion}
Using FWC, in Sec.~\ref{Sec: Metric in FWC}, we first calculated the Schwarzschild metric, Eq.~\eqref{eq: Schwarzschild metric}, as seen by the static observer located at the equator of the Earth, c.f., Eqs.~\eqref{eq: worldline observer} and~\eqref{eq: Metric in FWC}. This metric can be used for an accurate description of physical phenomena in a tube around the worldline of the observer, provided that the observed particles are sufficiently close. In our case, we expanded the metric up to second order in the spatial FWC normalized by the radius of the Earth, i.e., up to the order $\mathcal{O}(r^2/R_E^2)$. In contrast to Riemann-Normal coordinates~\cite{Nesterov1999,Kajari2009,Kajari2011}, the tetrad basis follows the worldline of the observer infinitely long. The general metric in FWC, Eq.~\eqref{eq: Metric in FWC (general)}, as well as the Schwarzschild metric in FWC, Eq.~\eqref{eq: Metric in FWC}, should be understood as a spatial expansion around the worldline of the observer, Eq.~\eqref{eq: worldline observer}. Note, that there is no other restriction on short times $\tau$ in the result of Sec.~\ref{Sec: Metric in FWC} than the restriction obtained by using the Schwarzschild metric, i.e., the experiments described should take place in a small time period compared to the time scale associated with the angular frequency of the Earth. When considering rotational effects, one could also apply our method to, e.g., the Kerr metric. Here, one would need to consider a co-rotating observer and make use of \textit{generalized} FWC~\cite{Kajari2009,Kajari2011,Llosa2017}. In that case, though, one should expect only semi-analytical results.

In Sec.~\ref{Sec: Generators of Symmetry Group}, while solving the Killing equations Eq.~\eqref{eq: Killing equation} for the metric  Eq.~\eqref{eq: Metric in FWC} of Sec.~\ref{Sec: Metric in FWC}, we expanded everything up to the second order in all four spacetime coordinates and to first order in $\epsilon=R_S/R_E$. Here, we restricted ourselves to a short time scale. In this way, we obtained the first order (in $\epsilon$) gravitationally corrected ten Poincaré symmetry generators, Eqs.~\eqref{eq: Generator time translations}--\eqref{eq: Boost generator}.

In Sec.~\ref{Sec: Generalized c.m. and relative Coordinates}, following the techniques of Osborn and Close~\cite{Osborn1968,Close1970}, we used these GR corrected Poincaré symmetry generators to calculate first order gravitational corrections for the c.m. coordinates and momentum, Eqs.~\eqref{eq:c.m. GR correction}. We then calculated the generalized relative coordinates, Eq.~\eqref{eq: relativistic relative Coordinates}, via the unitary transformation technique first presented by Liou~\cite{Liou1974}.

With these generalized c.m. and relative coordinates in hand as well as the Schwarzschild metric in FWC for a static observer, Eq.~\eqref{eq: Metric in FWC}, we then followed the works of Schwartz and Giulini~\cite{Schwartz2019,Schwartz2020b} to firstly obtain the total Lagrangian and ultimately the Hamiltonian describing a two-particle atom interacting with light in weakly curved spacetimes. In contrast to Schwartz and Giulini, when comparing to Refs.~\cite{Schwartz2019,Schwartz2020b}, we found a vanishing coupling between the internal dynamics and gravity and an additional term in the cross-coupling Hamiltonian, Eq.~\eqref{eq: Cross-term Hamiltonian}. Freezing out the radial motion and performing the unitary transformation, Eq.~\eqref{eq: unitary transformation cross-term},
this cross term vanishes and the remaining terms coupling internal and external dynamics can be interpreted as a state-dependent total mass, c.f., Eq.~\eqref{eq: c.m. Hamiltonian mass defect}.
Analyzing effects from possible transversal dynamics and their respective remaining cross terms, which would contradict the mass defect picture, remain an open question for future work.

Similar to the works of Sonnleitner et al. in flat spacetime~\cite{Sonnleitner2018} and their generalization to curved spacetime by Schwartz and Giulini~\cite{Schwartz2019,Schwartz2020b} we found the mass defect as coupling between the c.m. and relative degrees of freedom by a systematic derivation from GR principles rather than by \textit{ad hoc} considerations. However, in contrast to Schwartz and Giulini, we set our calculation on a realistic footing by using the metric as seen by a local observer and the GR corrected c.m. and relative coordinates. The total Hamiltonian, Eq.~\eqref{eq: total Hamiltonian}, as the final result of this manuscript can now be used for an accurate description of quantum optical experiments, e.g., atom interferometry.
The mass defect can be used for tests of the universality of free fall and the universality of gravitational redshift.
Hence, we provided the basis for a more detailed description of general relativistic effects and their measurement with quantum sensors such as clocks~\cite{Delva2018,Takamoto2020,Bothwell2021} and atom interferometers~\cite{Loriani2019,Roura2020,Ufrecht2020,DiPumpo2021,DiPumpo2022,DiPumpo2023,Janson2024}.
As the main goal of this manuscript was to derive the GR correction terms for the c.m. and relative coordinates in order to find the corresponding light-matter Hamiltonian in the presence of gravity for a local observer, we leave the practical applications, e.g., in the context of atom interferometry, open for future work.

\begin{acknowledgments}
We are grateful to W. P. Schleich for his stimulating input and continuing support.
We also thank F. Di Pumpo, A. Wolf, S. Böhringer, and A. Friedrich for many interesting and fruitful discussions.

This work was supported by the Science Sphere Quantum Science of Ulm University.
The QUANTUS and INTENTAS projects are supported by the German Space Agency at the German Aerospace Center (Deutsche Raumfahrtagentur im Deutschen Zentrum für Luft- und Raumfahrt, DLR) with funds provided by the Federal Ministry for Economic Affairs and Climate Action (Bundesministerium für Wirtschaft und Klimaschutz, BMWK) due to an enactment of the German Bundestag under Grant Nos. 50WM2450D (QUANTUS VI) and 50WM2178 (INTENTAS).
\end{acknowledgments}

\section*{Author Declarations}
\subsection*{Conflict of Interest Statement}
The authors have no conflicts to disclose.

\subsection*{Author Contributions }
\noindent{\sffamily\small\textbf{Gregor Janson}} Conceptualization (equal); Formal analysis (lead); Validation (equal); Investigation (lead); Methodology (equal); Visualization (lead); Writing - original draft (lead); Writing – review and editing (equal).
{\sffamily\small\textbf{Richard Lopp}} Conceptualization (equal); Formal analysis (support); Validation (equal); Investigation (support); Methodology (equal); Writing - original draft (Support); Writing – review and editing (equal); Supervision (lead).

\section*{Data Availability}

The data that support the findings of this study are available
within the article.

\appendix

\section{Maxwell equations in curved spacetime}\label{App: Maxwell equations}
In this section we want to solve Maxwell's equations for the Schwarzschild observer located on the equatorial plane. The corresponding metric is written down in Eq.~\eqref{eq: Metric in FWC}. We thus have to solve Maxwell's equations in curved spacetime that can be obtained by varying the action
\begin{align}
    S_\textsf{em} = \int\mathrm{d}t L_\textsf{em},
\end{align}
where $L_\textsf{em}$ is the electromagnetic Lagrangian, Eq.~\eqref{eq: em. Lagrangian}, with respect to the electromagnetic four-potential, $A_{\textsf{tot}\,\mu}$, leading to
\begin{align}\label{eq: Maxwell's equations (general)}
    \nabla_\mu F_\textsf{tot}^{\mu\nu} = -\frac{1}{\epsilon_0c^2}\frac{1}{\sqrt{-g}}j^\nu \Leftrightarrow \nabla^\mu F_{\textsf{tot}\,\mu\nu} = -\frac{1}{\epsilon_0c^2}\frac{1}{\sqrt{-g}}j_\nu,
\end{align}
where 
\begin{align}\label{eq: current density}
    j^\nu(t,\vec{r}) = \sum_{i=1}^{2}e_i\delta^{(3)}(\vec{r}-\vec{r}_i(t))\dot{\vec{r}}_i^\mu(t)
\end{align}
is the four-current density.
We can then use the Coulomb gauge
\begin{align}\label{eq: Coulomb gauge}
    \vec{\nabla}\cdot\vec{A}_{\textsf{tot}} = \partial_a A_{\textsf{tot}}^a = 0
\end{align}
and the Helmholtz decomposition of the vector potential $\vec{A}_\textsf{tot}$, i.e., a decomposition into a gradient of a scalar potential and a divergence-free part  $\vec{A}^\perp_\textsf{tot}$,
\begin{align}
    \vec{A}_\textsf{tot} = \vec{A}^\parallel_\textsf{tot} + \vec{A}^\perp_\textsf{tot}.
\end{align}
The Coulomb gauge, Eq.~\eqref{eq: Coulomb gauge}, requires $\vec{A}^\parallel_\textsf{tot.}=0$ and we are left with \mbox{$\vec{A}_\textsf{tot} = \vec{A}^\perp_\textsf{tot}$.} The covariant derivative of the field strength tensor can be easily derived and inserted into Maxwell's equations, Eq.~\eqref{eq: Maxwell's equations (general)}, which leads us to
\begin{subequations}\label{eq: Maxwell's equations tot.}
    \begin{align}
        &\Delta \phi_{\textsf{el,tot}} = -\frac{\rho}{\epsilon_0} + \frac{GM_E}{R_E^2c^2}\left(\frac{\rho}{\epsilon_0}x-3\partial_x\phi_\textsf{el,tot}+\partial_t A_{\textsf{tot}}^{\perp 1}\right)\\
        \begin{split}
        &(\Delta - c^{-2}\partial_t^2) \vec{A}_{\textsf{tot}}^\perp \\
        &= c^{-2}\left[\partial_t \vec{\nabla} \phi_\textsf{el,tot}-\frac{1}{\epsilon_0}\vec{j}+\frac{GM_E}{R_E^2} \left[\vec{e}_x\times\left(\vec{\nabla}\times\vec{A}_{\textsf{tot}}^\perp\right)\right]\right],
        \end{split}
    \end{align}
\end{subequations}
i.e., the Maxwell's equations for the electromagnetic scalar potential $\phi_\textsf{el,tot} = c A_\textsf{tot}^0$ and the vector potential $\vec{A}^\perp_\textsf{tot} = A_\textsf{tot}^{\perp a}$, where $\rho=c^{-1}j^0$ is the charge density and $\Delta=\delta^{ab}\partial_a\partial_b$ is the flat-spacetime Laplacian. We will now separate the internal and external parts of the electromagnetic fields, i.e.,
\begin{subequations}
    \begin{align}
        \phi_\textsf{el,tot} &= \phi_\textsf{el} + \phi_\textsf{el,ext}\\
        \vec{A}_\textsf{tot}^\perp &= \vec{\mathcal{A}}^\perp + \vec{A}^\perp
    \end{align}
\end{subequations}
and solve the corresponding Maxwell's equations separately.

\subsection{Internal part}
Firstly, let us consider the internal part of the electromagnetic fields and expand them in orders of $c^{-2}$, i.e.,
\begin{subequations}
    \begin{align}
        \phi_\textsf{el} &= \phi_\textsf{el}^{(0)} + c^{-2}\phi_\textsf{el}^{(2)} + \mathcal{O}(c^{-4})\\
        \vec{\mathcal{A}}^\perp &= \vec{\mathcal{A}}^{\perp (0)} + c^{-2}\vec{\mathcal{A}}^{\perp (2)} + \mathcal{O}(c^{-4}).
    \end{align}
\end{subequations}
We can then solve the internal Maxwell's equations order by order. We are only interested in solutions up to the order $c^{-2}$, so that we are left with
\begin{subequations}\label{eq: internal scalar potential equations}
    \begin{align}
        \Delta \phi_\textsf{el}^{(0)} &= -\frac{\rho}{\epsilon_0}\\
        \Delta \phi_\textsf{el}^{(2)} &= \frac{GM_E}{R_E^2}\left(\frac{\rho}{\epsilon_0}x-3\partial_x\phi_\textsf{el}^{(0)}+\partial_t \mathcal{A}^{\perp(0) 1}\right)
    \end{align}
\end{subequations}
for the internal scalar potential and
\begin{subequations}\label{eq: internal vector potential equations}
    \begin{align}
        (\Delta - c^{-2}\partial_t^2) \vec{\mathcal{A}}^{\perp (0)} &= 0\\
        \begin{split}
        (\Delta - c^{-2}\partial_t^2) \vec{\mathcal{A}}^{\perp (2)} &= \partial_t \vec{\nabla} \phi_\textsf{el}^{(0)}-\frac{1}{\epsilon_0}\vec{j}\\
        &\quad\quad+\frac{GM_E}{R_E^2} \left[\vec{e}_x\times\left(\vec{\nabla}\times\vec{\mathcal{A}}^{\perp (0)}\right)\right]
        \end{split}
    \end{align}
\end{subequations}
for the internal vector potential. The lowest order equations for the internal scalar and vector potentials can be easily solved. Eq.~\eqref{eq: internal scalar potential equations} is the non-gravitational Poisson equation in lowest order that is solved by
\begin{align}\label{eq: lowest order internal scalar potential}
    \phi_\textsf{el}^{(0)}(\vec{r}) = \frac{1}{4\pi\epsilon_0}\int\mathrm{d}^3r' \frac{\rho(\vec{r}')}{\vert\vec{r}-\vec{r}'\vert} = \frac{1}{4\pi\epsilon_0}\left[\frac{e_1}{\vert\vec{r}-\vec{r}_1\vert}+\frac{e_2}{\vert\vec{r}-\vec{r}_2\vert}\right].
\end{align}
Moreover, the internal fields should only describe effects originated from the two particles itself. That is why there should not be radiative terms for the lowest order internal vector potential. We thus have
\begin{align}\label{eq: lowest order internal vector potential}
    \vec{\mathcal{A}}^{\perp (0)} = 0.
\end{align}
The lowest order solutions can then be plugged into the Eqs.~\eqref{eq: internal scalar potential equations} and~\eqref{eq: internal vector potential equations} to solve the next higher order. For the internal scalar potential we find
\begin{align}\label{eq: second order internal scalar potential}
    \phi_\textsf{el}^{(2)}(\vec{r}) = -\frac{GM_E}{4\pi R_E^2}\int\mathrm{d}^3r' \left(\frac{\rho(\vec{r}')}{\epsilon_0}x'-3\partial_{x'}\phi_\textsf{el.}^{(0)}(\vec{r}')\right)/\vert\vec{r}-\vec{r}'\vert
\end{align}
and the equation for $c^{-2}\vec{\mathcal{A}}^{\perp (2)}$ is the non-gravitational wave equation that was already solved by Sonnleitner et al. in Appendix A of Ref.~\cite{Sonnleitner2018}:
\begin{align}
    \begin{split}
        &\vec{\mathcal{A}}^{\perp}(\vec{r}) = \vec{\mathcal{A}}^{\perp}_\textsf{ng}(\vec{r}) + \mathcal{O}(c^{-4}) \\
        = &\frac{1}{8\pi\epsilon_0c^2}\sum_{i=1}^2 e_i\left(\frac{\dot{\vec{r}}_i}{\vert\vec{r}-\vec{r}_i\vert}+\frac{(\vec{r}-\vec{r}_i)[\dot{\vec{r}}_i\cdot(\vec{r}-\vec{r}_i)]}{\vert\vec{r}-\vec{r}_i\vert^3}\right) + \mathcal{O}(c^{-4}).
    \end{split}
\end{align}
We do not need to calculate the integral in Eq.~\eqref{eq: second order internal scalar potential}. Instead one can show that
\begin{align}
\begin{split}
    \int\mathrm{d}^3r\; j^0(\vec{r}) \mathcal{A}_0 &= -\int\mathrm{d}^3r\; \rho(\vec{r}) \phi_\textsf{el}(\vec{r})\left(1+2\frac{G M_E}{R_E^2 c^2}x\right)\\
    &= -\frac{e_1 e_2}{2\pi\epsilon_0\vert\vec{r}_1 - \vec{r_2}\vert}\left[1+\frac{1}{2}\frac{G M_E}{R_E^2 c^2} (x_1 + x_2)\right],
\end{split}
\end{align}
which is the corresponding term in the electromagnetic Lagrangian.

\subsection{External part}
Analogously to the previous section we can expand the external electromagnetic fields in orders of $c^{-2}$, i.e.,
\begin{subequations}
    \begin{align}
        \phi_\textsf{el,ext} &= \phi_\textsf{el,ext}^{(0)} + c^{-2}\phi_\textsf{el,ext}^{(2)} + \mathcal{O}(c^{-4})\\
        \vec{A}^\perp &= \vec{A}^{\perp (0)} + c^{-2}\vec{A}^{\perp (2)} + \mathcal{O}(c^{-4}).
    \end{align}
\end{subequations}
However, we now have a slightly different situation. We assume that there are no external charges, i.e., $\rho=j^a=0$. Therefore, the external Maxwell's equations read
\begin{subequations}\label{eq: external scalar potential equations}
   \begin{align}
        \Delta\phi_\textsf{el,ext}^{(0)} &= 0\\
        \Delta\phi_\textsf{el,ext}^{(2)} &= \frac{GM_E}{R_E^2}\left(-3\partial_x \phi_\textsf{el,ext}^{(0)} + \partial_t A^{\perp  1 (0)}\right)
   \end{align}
\end{subequations}
for the external scalar potential and
\begin{subequations}\label{eq: external vector potential equations}
    \begin{align}
        (\Delta - c^{-2}\partial_t^2) \vec{A}^{\perp (0)} &= 0\\
        (\Delta - c^{-2}\partial_t^2) \vec{A}^{\perp (2)} &= \partial_t \vec{\nabla} \phi_\textsf{el,ext}^{(0)}+\frac{GM_E}{R_E^2}\left[\vec{e}_x\times\left(\vec{\nabla}\times\vec{A}^{\perp (0)}\right)\right]
    \end{align}
\end{subequations}
for the external vector potential. We do not need to solve these equations explicitly. We only need to gain some knowledge about the respective order in $c^{-1}$. Eq.~\eqref{eq: external scalar potential equations} gives immediately
\begin{align}
    \phi_\textsf{el,ext} = \mathcal{O}(c^{-2}).
\end{align}
In contrast to the internal vector potential, we now allow radiative solutions for the external part. The wave equation, Eq.~\eqref{eq: external vector potential equations}, for the lowest order of the external vector potential gives us
\begin{align}
    \partial_a\vec{A}^\perp = \mathcal{O}(c^{-1}),
\end{align}
where we have used $\partial_t\vec{A}^\perp = \mathcal{O}(c^0)$ corresponding to the external electric field.

\bibliography{main.bib}

\end{document}